\documentclass[aps,prd,amsmath,amsfonts,letterpaper,nofootinbib,notitlepage,superscriptaddress]{revtex4}

\usepackage{graphicx,hyperref,mathrsfs}

\begin{document}

%%%%%%%%%%%%%%%%%%%%%%%%%%%%%%%%%%%%%%%%%%%%%%%%%%%%%%%%%%%%%%
\title{Non-Gaussian features of primordial magnetic fields in power-law inflation}
\author{Leonardo Motta}\email{leonardo.motta@dartmouth.edu}
\author{Robert R. Caldwell}
\affiliation{Department of Physics and Astronomy, Dartmouth College, HB 6127 Wilder Lab, Hanover, NH 03755, USA}

\date{\today}

%%%%%%%%%%%%%%%%%%%%%%%%%%%%%%%%%%%%%%%%%%%%%%%%%%%%%%%%%%%%%%
\begin{abstract} 
We show that a conformal-invariance violating coupling of the inflaton to electromagnetism produces a cross correlation between curvature fluctuations and a spectrum of primordial magnetic fields. According to this model, in the case of power-law inflation, a primordial magnetic field is generated with a nearly flat power spectrum and rms amplitude ranging from nG to pG. We study the cross correlation, a three-point function of the curvature perturbation and two powers of the magnetic field, in real and momentum space. The cross-correlation coefficient, a dimensionless ratio of the three-point function with the curvature perturbation and magnetic field power spectra, can be several orders of magnitude larger than expected as based on the amplitude of scalar metric perturbations from inflation. In momentum space, the cross-correlation peaks for flattened triangle configurations, and is three orders of magnitude larger than the squeezed triangle configuration. These results suggest likely methods for distinguishing the observational signatures of the model.
\end{abstract}

\maketitle

%%%%%%%%%%%%%%%%%%%%%%%%%%%%%%%%%%%%%%%%%%%%%%%%%%%%%%%%%%%%%%
\section{\label{sec:intro}Introduction}

The simplest models of inflation have proven remarkably successful at solving an array of cosmological problems whilst bringing to order a vast catalog of observations. However, little is known about the Lagrangian during inflation. Non-gaussian correlations of cosmological observables carry information about the details of the physics during inflation. In this article, we study what sort of dominant non-gaussian features should be expected in the statistics of primordial magnetic fields arising from a coupling between the inflaton and electrodynamics.

The Lagrangian we examine is motivated by the problem of galactic magnetism. It is observed that galaxies are permeated with magnetic fields of order 100 to 1 $\mu\text{G}$. This can be seen in great detail in the recent study of magnetic fields in M51 by Ref.~\cite{Fletcher:2010wt}, for example. The current understanding of the formation and evolution of these fields is that a dynamo mechanism, combining galactic rotation with helical turbulence, amplifies a seed magnetic field with strength larger than $10^{-20}\,\text{G}$ at the time of the galactic disk formation \cite{Widrow:2002ud,Beck:2005cm,Brandenburg:2004jv}. The seed field must have been more or less homogeneous on the scale of the galactic disk when galaxies started to form, implying a comoving coherence scale of around a Mpc. Such a large scale has been inside the horizon since a redshift $z\sim10^{6}$, however the physics from the time of Big Bang Nucleosynthesis at $z\sim 10^{8}$ to the present is well-established and, in the standard cosmology, no large-scale magnetic field of significant strength is known to have been generated inside the horizon since then. In fact, the largest seed field produced from the nonlinear evolution of subhorizon primordial density perturbations, while remaining consistent with the physics of recombination, is only $10^{-29}\;\text{G}$ on Mpc scales at the present time~\cite{Matarrese:2004kq}, which is far too small to seed the dynamo mechanism. Logically, then, we may ask whether a superhorizon magnetic field produced during inflation could have provided the seed field. However, quantum electrodynamics is conformally invariant, so that the amplification mechanism that produces density pertubations during inflation leaves an uninterestingly small magnetic field energy density. To explain such a seed field, new physics would be required during inflation to break conformal invariance~\cite{Turner:1987bw}.

In a previous article \cite{Caldwell:2011ra}, a toy model for such new physics was investigated in which a spectator field in de Sitter spacetime couples to electromagnetism. In the absence of gravity, it was shown that a scale-free magnetic field spectrum with rms amplitude $\sim 0.1$~nG at Mpc scales may be achieved. The central result of that investigation was the cross correlation between the scalar field and magnetic fields, showing that the dimensionless amplitude, measured in units of the power spectra, can grow as large as $\sim 500 H_I/M$, where $H_I$ is the inflationary Hubble parameter and $M$ is the effective mass scale of the coupling.

In this article, we consider a more realistic scenario, coupling the inflaton of power-law inflation to electromagnetism with a dilaton-type interaction. Many of the results from the earlier investigation will be shown to carry over, including the result of a nearly scale-invariant magnetic field spectrum with rms amplitude ranging from $10^{-3}-1$~nG at Mpc scales.  However the present work concentrates on the issue of the non-gaussian corrections to the statistics of the cosmic magnetic field \cite{Brown:2005kr}. These are contained in terms such as the one-loop correction to the magnetic field power spectrum,  four-point function, or the cross correlation with metric perturbations. The first two involve at least two insertions of the interaction Hamiltonian in the in-in formalism perturbative series, hence are expected to be subdominant in comparison to the latter. Since tensor perturbations are harder to detect than scalar metric perturbations, the most important non-gaussianity in magnetic field spectra arises in the cross correlation between the curvature fluctuation $\mathcal{R}$ and the magnetic field energy density, in the form $\langle \mathcal{R} B^2\rangle$ which we evaluate in this paper. This cross correlation, which we naively expect to have amplitude $10^{-5}$ to match the amplitude of scalar metric perturbations, may be probed directly in experiments that are sensitive to the statistics of the large-scale, initial conditions in the mass density and magnetic fields, such as cosmic microwave background temperature-polarization correlations and the correlation of all-sky Faraday rotation measures with galactic density~\cite{Stasyszyn:2010kp}. 

The outline of the paper is as follows. In Sec.~\ref{sec:themodel} we introduce our model. In Sec.~\ref{sec:threepointfunc} we present the calculation of the cross correlation, and we study its properties in Sec.~\ref{sec:bispectrum}. We conclude in Sec.~\ref{sec:conclusions}. In regard to notation, a prime indicates a derivative with respect to conformal time, $' = d/d\eta$, an overdot indicates a derivative with respect to cosmic time, $\dot{} = d/d t$, and $dt/d\eta = a$ where $a$ is the expansion scale factor. When necessary, an asterisk subscript denotes a quantity evaluated outside the horizon.

%%%%%%%%%%%%%%%%%%%%%%%%%%%%%%%%%%%%%%%%%%%%%%%%%%%%%%%%%%%%%%
\section{\label{sec:themodel}Relic Magnetic Fields from Inflaton}

%%%%%%%%%%%%%%%%%%%%%%%%%%%%%%%%%%%%%%%%%%%%%%%%%%%%%%%%%%%%%%
\subsection{Amplification mechanisms}

In an effective field theory of inflation, it is possible to couple the inflaton $\varphi$ to electrodynamics at the lowest dimensionality in general form with an action 
\begin{equation}\label{eq:actionWFF}
S=\int d^4 x \, \sqrt{-g} \; \Biggl[ -\frac{M_\text{Pl}^2}{16\pi} R - \frac{1}{2} g^{\mu\nu}\partial_\mu\varphi\partial_\nu\varphi -V(\varphi) -\frac{1}{4} g^{\mu\lambda}g^{\nu\sigma}\, W(\varphi) F_{\mu\nu} F_{\lambda\sigma}\Biggr]
\end{equation}
where $W$ is a dimensionless scalar built out of $\varphi$ and its derivatives. We envisage an ultraviolet completion of the effective Lagrangian in Eq.~(\ref{eq:actionWFF}) where integrating out heavy fields leads to the inflaton potential $V(\varphi)$ and the coupling $W(\varphi)$. In the very low energy limit, we require $W$ to be constant, so that quantum fluctuations of the electromagnetic field are conformally invariant, and there is no coupling between electrodynamics and the inflaton. However, conformal invariance is broken when $W$ evolves in time or varies in space. 

In a previous work, we showed that by driving $W$ with a spectator field during inflation, then the evolution $W \propto a^{4}$ produces a relic cosmological magnetic field as large as $10^{-10}\,\text{G}$ on Mpc scales at the present time \cite{Caldwell:2011ra}. A similar approach was previously studied in Refs.~\cite{Ratra:1991bn,Gasperini:1995dh,Bamba:2003av}. Our function $W$ is a special case of the function denoted $I^2$ in the general case studies of Refs.~\cite{Bamba:2006ga,Demozzi:2009fu}. Our plan in this work is to show that once an amplification mechanism has been selected and the power spectrum of magnetic fields is used to fix the parameters of the Lagrangian in Eq.~(\ref{eq:actionWFF}), one can go further and predict the structure of the cross correlation of scalar metric fluctuations and magnetic fields on cosmological scales.

As a concrete effective theory, we consider slow-roll inflation with the potential~\cite{Abbott:1984fp,Lucchin:1984yf,Lucchin:1985wy,Lyth:1991bc}
\begin{equation}\label{eq:exponentialpotential}
V(\varphi) = M^4 \exp(-\varphi/\Lambda)\, 
\end{equation}
where $M$ and $\Lambda$ are two mass scales. In a theory where $\exp(\varphi)$ appears in the Lagrangian, it is natural to take as the coupling function
\begin{equation}\label{eq:Weg}
W(\varphi) = \exp\left(\frac{n\varphi}{\epsilon \Lambda}\right)
\end{equation}
where $\epsilon = -\dot{H}/H^2 = 1/(16\pi G \Lambda^2)=M_\text{Pl}^2/(16 \pi \Lambda^2)$, $n$ is a constant parameter, and we indicate the Planck mass as $M_\text{Pl} = 1/\sqrt{G}$. The appeal of the single-field inflaton potential of Eq.~(\ref{eq:exponentialpotential}) is that it admits a closed form solution for the scale factor and the evolution of the fields in time without any further approximation in the size of $\epsilon$, namely $H=1/\epsilon t$, $a(t) = a_I (t/t_I)^{1/\epsilon}$, or $a(\eta) = a_I (\eta_I/\eta)^{1/(1-\epsilon)}$ in terms of conformal time, where the subscript $I$ indicates values at the end of inflation, and 
\begin{equation}
{\varphi}(t) = \Lambda \log\left(\frac{8 \pi \epsilon^2 M^4t^2}{(3-\epsilon)M_\text{Pl}^2}\right).
\end{equation}
Therefore, the coupling function is 
\begin{equation}\label{eq:Wa2}
W({\varphi})=W_I (a/a_I)^{2n}\, 
\end{equation}
where $W_I$ is the coupling at the end of inflation. 

For arbitrary inflation models, it is always possible to start with a generally covariant gravitational form like Eq.~(\ref{eq:Weg}) and arrive at the specific slow-roll behavior of Eq.~(\ref{eq:Wa2}). This is because the scale factor for generic single field slow-roll evolves according to
\begin{equation}
a(t)\simeq \exp\left[-\int^{\varphi(t)} \left(\frac{V(\varphi)}{M_\text{Pl}^2 V'(\varphi)} \right) d\varphi \right]
\end{equation}
and the flatness condition requires that $V(\varphi)/V'(\varphi)$ is approximately a constant as a function of $\varphi$, hence $a(t)$ will be a power of $\exp(\varphi/v)$. Therefore, $W$ will be proportional to $a$ if it is a power of $\exp(\varphi/v)$. Then, as reheating takes place, the inflaton field value and its derivatives approach zero, so that $W\rightarrow 1$. Consequently, there is no modification of electrodynamics at later stages (or lower energies).

%%%%%%%%%%%%%%%%%%%%%%%%%%%%%%%%%%%%%%%%%%%%%%%%%%%%%%%%%%%%%%
\subsection{Quantum fluctuations of the electromagnetic field}

The action Eq.~(\ref{eq:actionWFF}) of the free electromagnetic field is
\begin{equation}
S_A = -\int d^4 x \sqrt{-g} \,W(\varphi)\frac{1}{4} F_{\mu\nu}F^{\mu\nu} 
= \int d\eta \,d^3 x \; W_I \left({a(\eta)}/{a(\eta_I)}\right)^{2n} \left( \frac{1}{2} F_{0 i}^2 - \frac{1}{4} F_{ij}^2\right),
\end{equation}
where above and hereafter, Latin indices $i,\,j,\, ...$ indicate components of the comoving Cartesian coordinate system.
The quantum field theory will avoid ambiguities if we choose a complete gauge specification, thus we proceed in the Coulomb gauge defined by $\partial_i A_i = 0$ in these coordinates. Gauss' law is then satisfied with $A_0 = 0$, and the action further simplifies to
\begin{equation}
S_A = \int d\eta \, d^3 x \; W_I \left({a(\eta)}/{a(\eta_I)}\right)^{2n}  \left(\frac{1}{2} A'^2_i - \frac{1}{2} (\partial_i A_j )^2 \right)\, .
\label{eqn:actiona}
\end{equation}
The canonical field conjugate to $A_i$ is $\Pi_i(\mathbf{x},\eta)= \delta S / \delta A'_i(\mathbf{x},\eta) =W_I (a/a_I)^{2n} A'_i(\mathbf{x},\eta)$, thus the canonical commutation relation $\left[A_i(\mathbf{x},\eta),\Pi_j(\mathbf{y},\eta)\right] = i \delta_{ij} \delta^3(\mathbf{x}-\mathbf{y})$ is here equivalent to
\begin{equation}\label{eq:AA'commutation}
\left[ A_i (\mathbf{x}, \eta), A'_j(\mathbf{y}, \eta)\right] = \frac{i}{W_I} (a_I/a)^{2n} \delta_{ij} \delta^3(\mathbf{x}-\mathbf{y})\, .
\end{equation}
We perform the usual expansion 
\begin{equation}
A_i(\mathbf{x}, \eta) = \sum_\sigma \int \frac{d^3 k}{(2 \pi)^3} \left[ v_k(\eta) e^{i\mathbf{k}\cdot\mathbf{x}} \mathbf{e}^{(\sigma)}_i(\hat{\mathbf{k}}) \beta(\mathbf{k}, \sigma) + \text{H.c.} \right] 
\end{equation}
where $\beta(\mathbf{k},\sigma)$ and $\beta(\mathbf{k},\sigma)^\dagger$ are the annihilation and creation operators, which satisfy $\left[ \beta(\mathbf{k},\sigma), \beta(\mathbf{k'},\sigma')^\dagger\right]=(2 \pi)^3 \delta_{\sigma\sigma'} \delta^3(\mathbf{k}-\mathbf{k}')$. Moreover, the polarization vectors $\mathbf{e}^{(\sigma)}_i(\hat{\mathbf{k}})$ 
satisfy Eq.~(\ref{eq:AA'commutation}) if
\begin{equation}
\sum_\sigma e^{(\sigma)}_i(\hat{\mathbf{k}})e^{(\sigma)*}_j(\hat{\mathbf{k}})=\delta_{ij}-\hat{k}_i\hat{k}_j\, .
\end{equation}
Finally,
\begin{equation}
v_k(\eta) v'^*_k(\eta) - v^*_k(\eta) v'_k(\eta) = \frac{i}{W_I}{ \left(a_I/a\right)^{2 n}}  \,  \label{eq:Wronskianv}
\end{equation}
gives the Wronskian condition on the mode function $v_k(\eta)$. 

The equation of motion for the gauge field mode function following from Eq.~(\ref{eqn:actiona}) is given by
\begin{equation}\label{eq:eulerv}
v''_k + 2n \frac{a'}{a} v'_k + k^2 v_k = 0\, .
\end{equation}
For successful inflation, all relevant cosmological modes must have been deep inside the horizon at an early enough time:
\begin{equation}\label{eq:vin}
v_k \sim \frac{e^{-ik\eta}}{\sqrt{2 k W_I} (a/a_I)^n }\, , \qquad -k\eta\rightarrow \infty\, ,
\end{equation}
which is consistent with the Wronskian condition, Eq.~(\ref{eq:Wronskianv}), and fixes the solution of Eq.~(\ref{eq:eulerv}) we seek:
\begin{eqnarray}
v_k(\eta) &=&   \left( \frac{-\pi \eta_I }{4 W_I} \right)^{1/2} e^{i\pi\alpha/2+i\pi/4}   (\eta/\eta_I)^\alpha H^{(1)}_\alpha (-k\eta) \\
\label{eq:vtau}
&=&  v_{k*} u_\alpha(-k \eta) \\
u_\alpha(x) &\equiv & \frac{i \pi x^\alpha}{2^\alpha \Gamma(\alpha)}H^{(1)}_\alpha(x)
\end{eqnarray}
where $\alpha=1/2+n/(1-\epsilon)$. 
The constant coefficient $v_{k*}$ is the value of the mode function outside the horizon,
$v_{k}(\eta) = v^*_k$ for $-k\eta \ll 1$ where
\begin{equation}
v_{k*}=-i\frac{2^{\alpha-1}\Gamma(\alpha)}{\pi^{1/2}W_I^{1/2}} \frac{(-k\eta_I)^{1/2-\alpha}}{k^{1/2}}e^{i\pi\alpha/2+i\pi/4}\, .
\label{eq:vout}
\end{equation}
Thus, Mpc modes of the electromagnetic field are stretched beyond the Hubble scale during inflation, where they are frozen until re-entering in the present epoch. The result of Eq.~(\ref{eq:vout}) will prove useful in determining the size of the magnetic field at the present time.

%%%%%%%%%%%%%%%%%%%%%%%%%%%%%%%%%%%%%%%%%%%%%%%%%%%%%%%%%%%%%%
\subsection{\label{sec:Bpowerspec}Magnetic field power spectrum}

The magnetic field two-point correction function is given by
\begin{eqnarray}
\langle B_i (\mathbf{x},\eta) B^i(\mathbf{y},\eta) \rangle & =&  \frac{1}{a^4} \left( \delta_{ij} \frac{\partial^2}{\partial x^k \partial y^k} - \frac{\partial^2}{\partial x^j \partial y^i}\right) \langle A_i(\vec x,\eta) A_j(\vec y,\eta) \rangle  \cr
&=& \int \frac{d^3 k}{(2 \pi)^3} e^{i\vec k \cdot (\vec x- \vec y)} P_B(k) \cr
P_B(k) &=& 2 \frac{k^2 |v_k|^2}{a^4} =
\frac{\pi}{2} \frac{H(\eta)^4(1-\epsilon)^4}{W(\eta) k^3} (-k \eta)^5 H^{(1)}_{\alpha}(-k \eta) H^{(2)}_{\alpha}(-k \eta).
\end{eqnarray}
The above spectrum is identical to the result obtained in our previous work, Ref.~\cite{Caldwell:2011ra}, upon taking the limit $\epsilon \to 0$. The power spectrum for Fourier modes outside the horizon at the end of inflation,
when  $|k\eta_I | \ll 1$, is given by
\begin{eqnarray}
\Delta^2_B(k) \equiv \frac{k^3}{2 \pi^2} P_B &=& \left(\frac{2}{\pi}\right)^3 \frac{\Gamma(\frac{5-n_B}{2})^2}{2^{n_B} W_I} (1-\epsilon)^4 H_I^4 (-k \eta_I)^{n_B} \\
& = &  
 \left(\frac{2}{\pi}\right)^3 \frac{\Gamma(\frac{5-n_B}{2})^2}{2^{n_B} W_I} (1-\epsilon)^{4-n_B} H_*^4 \left( \frac{a_*}{a_I}\right)^{4 \epsilon + n_B(1-\epsilon)}  \left(\frac{k}{a_* H_*}\right)^{n_B}.
\end{eqnarray}
where $n_B = 4 - 2 n/(1-\epsilon)$ for $n>0$. In the second line above, we express the power spectrum at the end of inflation in terms of the scale factor and Hubble parameter at horizon cross, namely $k/a_*=H_*$. We can see that $n_B=0$, or $n= 2(1-\epsilon)$, gives a flat spectrum. (A flat spectrum can also be achieved  for $n=-3(1-\epsilon)$ but this solution is not viable, as discussed in the next section.) After inflation, when $W(\varphi)$ becomes a constant, conformal invariance is restored and the magnetic field amplitude will decay as $a^{-2}$. Under the assumption of instantaneous reheating, the spectrum today will be $(a_I/a_0)^4$ times the above result. 

To obtain a numerical value of the power spectrum, we start by fixing $\epsilon \simeq 0.016$ and $H_{*} \simeq 10^{14}$~GeV for Mpc modes, both values of which are consistent with WMAP. Under the assumption that the Hubble parameter does not change from its horizon cross value by many orders of magnitude during reheating,  one may write $3 H_*^2/8 \pi G = \pi^2 g_I T_I^4/30$ where $g_I$ and $T_I$ are the effective degrees of freedom and temperature of the radiation fluid at the end of inflation, and we assume adiabatic evolution thereafter. For this simplistic model, we use $g_I \sim 10^3$ to take into account possible degrees of freedom beyond the Standard Model. As a consequence, $T_I \simeq 3 \times 10^{15}$GeV and $H_I \simeq 5 \times 10^{13}$~GeV, corresponding to a redshift for the end of inflation of $z_I \simeq 10^{29}$. (Note that in Ref.~\cite{Caldwell:2011ra} we used $H_I= 10^{14}$~GeV and $z_I  = 10^{28}$.) With the aid of the conversion factor $({\rm Gauss})^2/8\pi = 1.91 \times 10^{-40}{\rm GeV}^4$, the energy density per logarithmic frequency interval of magnetic fields at the present time is
\begin{eqnarray}
\frac{d}{d \ln k}\langle B^2 \rangle = \Delta^2_B(k) & = &  
 \left(\frac{2}{\pi}\right)^3 \frac{\Gamma(\frac{5-n_B}{2})^2}{2^{n_B} W_I} (1-\epsilon)^{4-n_B} H_*^4 \left( \frac{a_*}{a_I}\right)^{4 \epsilon + n_B(1-\epsilon)}    \left( \frac{a_I}{a_0}\right)^{4}    \left(\frac{k}{a_* H_*}\right)^{n_B} \\
& \simeq &  10^{-22.8-22.5 n_B} \frac{\Gamma(({5-n_B})/{2})^2}{W_I\Gamma(5/2)^2} \left( \frac{k}{\rm Mpc^{-1}}\right)^{n_B}\, {\rm G}^2.
\label{eqn:Bpower}
\end{eqnarray}
If $n_B = 0$ and $W_I\simeq 1$ then the field strength is roughly \mbox{$10^{-12}\,\text{G}$} on Mpc scales,  whilst a slightly blue tilt as great as $n_B \sim -0.2$ yields $B_{\rm rms} \simeq 10^{-9}$~G, which in both cases may be sufficient to explain the observed astrophysical and cosmological magnetic fields.

%%%%%%%%%%%%%%%%%%%%%%%%%%%%%%%%%%%%%%%%%%%%%%%%%%%%%%%%%%%%%%
\subsection{\label{sec:Bbackreaction}Backreaction}

Self-consistency of this scheme for the generation of a primordial magnetic field requires that the amplification mechanism does not generate an energy density of gauge fields as large as the inflaton energy density and thereby spoil inflation. This is a real concern, since the $n=-3$ case, while corresponding to $n_B=0$, leads to an overproduction of electric field energy density. This issue was first analyzed in detail by \cite{Demozzi:2009fu}. As shown in our earlier work \cite{Caldwell:2011ra}, the energy density of the electromagnetic field as measured in the cosmic rest frame is
\begin{eqnarray}
\rho_{EB} &=& \frac{W(\eta)}{2 a(\eta)^4} \langle A'_i A'_i + (\partial_i A_j)(\partial_i A_j)\rangle \\
&=& \frac{W(\eta)}{2 \pi^2 a(\eta)^4}\int k^2 \, dk \, \left( |v_k'|^2 + k^2 |v_k|^2 \right) \\
\Omega_{EB} &=& \frac{1}{3}GH_I^2(1-\epsilon)^4 \int_{e^{-N_I}}^1 dx\, x^4 \,\left( |H^{(1)}_{\alpha}(x)|^2 + |H^{(1)}_{\alpha-1}(x)|^2 \right)\\
&=& G H_I^2 \times
\begin{cases}
{\cal O}(1), & {\rm for} \quad  |n|\le 2 \cr
{\cal O}(1) \times e^{2N_I(|n|-2)}, & {\rm for}\quad |n| > 2
\end{cases}
\end{eqnarray}
where $N_I$ is the number of e-foldings of inflation. Using $N_I=70$ as a fiducial value, we derive the limit $|n| \lesssim 2.1$.
Using $n_B = 4 - 2 n/(1-\epsilon)$ for $n>0$, this translates into the bound $-0.2 < n_B < 4$ within which $\Omega_{EB} < 0.01$, such that the energy density of quantum electrodynamics amplified by the inflaton will be small. This leaves the $n_B=0$ case as a potentially viable scenario for the creation of a scale-free spectrum of primordial magnetic fields.

%%%%%%%%%%%%%%%%%%%%%%%%%%%%%%%%%%%%%%%%%%%%%%%%%%%%%%%%%%%%%%
\subsection{Charge energy density contribution}

There is a further issue to be addressed in a successful model of primordial magnetogenesis, which was raised in Ref.~\cite{Demozzi:2009fu} regarding the effect of the electromagnetic current. To illustrate the issue, suppose that we consider the charged fields at the relevant scale $H_*$ to be complex scalars $\sigma_n$:
\begin{equation}\label{eq:lagrangianwmatter1}
S = \int d^4x \, \sqrt{-g}\left[-\frac{1}{4}W(\varphi)F^{\mu\nu}F_{\mu\nu} - J^\mu A_\mu - \sum_{n}g^{\mu\nu}\partial_\mu\sigma^{*}_n\partial_\nu\sigma_n  \right]
\end{equation}
and $J^\mu$ is the gauge-invariant covariantly-conserved current associated with the matter fields $\sigma_n$. Based on the discussion of the previous section, Sec.~\ref{sec:Bbackreaction}, we would like to take $n\sim 2$ which means $W\propto a^{4}$. If the ratio of Coulomb energy to electromagnetic kinetic energy is of order one at the end of inflation then it is of order $e^{4 N_I}$ at early times, where $N_I$ is the number of e-foldings, therefore breaking down perturbation theory. In other words, the electromagnetic coupling is proportionately large and therefore strongly coupled at the beginning of inflation. This would appear to be fatal for this model, since we could no longer justify a free-field, perturbative treatment of electromagnetism. Equivalently, if we start with $W$ of order one at the beginning of inflation in Eq.~(\ref{eq:lagrangianwmatter1}), then at the end of inflation the kinetic energy of the gauge fields would acquire a factor $e^{4N_I}$. If at this point we redefine the gauge field to canonical normalization, by scaling $A_\mu$ by a factor of $e^{-2N_I}$, then the classical electric charge becomes $e^{-2N_I} e_0$, where $e_0$ is the electric charge at the beginning of inflation. Since one expects $N_I \sim 65-70$ at least, the electric charge at the end of inflation is at least 130 orders of magnitude smaller than one. This is to be taken as the charge at the scale of reheating. If the beta function of the electric charge at these scales is negative, then its value may be brought to larger values at zero energy, but this would require an amplification of a similar number of orders of magnitude.

The problem of the strong coupling can be fixed if the effective field theory at the relevant scales breaks gauge invariance as follows:
\begin{equation}\label{eq:ActionWithCurrent}
S = \int d^4x \, \sqrt{-g}\,  \left[-\frac{1}{4} W(\varphi)\, F^{\mu\nu}F_{\mu\nu} - W(\varphi) J^\mu A_\mu - \sum_{n}g^{\mu\nu}\partial_\mu\sigma^{*}_n\partial_\nu\sigma_n \right]\, .
\end{equation}
With the convention $W=W_I(a/a_I)^{2n}$ and $W_I \sim 1$,  the action at the end of inflation ($a=a_I$) reads more explicitly
\begin{equation}
S = \int d^3 x \, d\eta \, \left[ -\frac{1}{4}\eta^{\mu\rho}\eta^{\nu\kappa}F_{\mu\nu}F_{\rho\kappa} - \sum_n a_I^2\left( i e_0 \sigma_n^*\partial_\mu\sigma_n - \text{H.c.}\right)\eta^{\mu\nu} A_\nu - \sum_n a_I^{2} \eta^{\mu\nu}\partial_\mu\sigma^*\partial_\nu\sigma\right]
\end{equation}
where $e_0$ is the electric charge during inflation. In this case, ignoring any running of the electric charge from reheating to zero momentum transfer, one can simply take $e_0$ to be the renormalized charge at zero momentum. There is no longer any problem at early times since both the current contribution and the kinetic energy of gauge fields go to zero as $a\rightarrow 0$ at the same rate. As pointed out in Ref.~\cite{Barnaby:2012tk} this would not succeed if gauge invariance is preserved.

In this work we will concentrate on the possible non-gaussian signature of the amplification mechanism and so we do not provide any rationale for writing a Lagrangian like Eq.~(\ref{eq:ActionWithCurrent}), which would have to arise from the UV completion. However, we point out that in theories with extra dimensions, gauge invariance can be violated in the 3+1 brane in a controllable way at high energies as a consequence of charges leaking to the extra dimensions, even if gauge invariance is maintained in the full space-time. Consider e.g. the discussion of this issue in Ref.~\cite{Dubovsky:2000av}. In this latter reference, it is shown how charged particles leaking to the extra dimensions produce electromagnetic waves in the 3+1 brane. It would be interesting to analyze whether this effect in brane-world models may provide a physical mechanism for the amplification factor $W$.

%%%%%%%%%%%%%%%%%%%%%%%%%%%%%%%%%%%%%%%%%%%%%%%%%%%%%%%%%%%%%%
\subsection{Curvature perturbations}

Quantum fluctuations of the inflaton field create inhomogeneities that may be treated in terms of perturbations of the spatial curvature, ${\cal R}$. A suitable set of the coordinates of the perturbed Robertson-Walker spacetime use the Arnowitt-Deser-Misner parametrization and consider only the linear scalar perturbation in the gauge in which the inflaton perturbations are set to zero $\delta\varphi = 0$ \cite{Maldacena:2002vr}. The metric to linear order is then given by
\begin{align}\label{eq:ADMmetric}
g_{00} = -N^2 + g_{ij}N^i N^j && g_{0i} = g_{ij} N^j && g_{ij} = a^2 (1 +  2 \mathcal{R}) \delta_{ij} \\
g^{00} = -\frac{1}{N^2} && g^{0i}= \frac{N^i}{N^2} && g^{ij} = \frac{1-2\mathcal{R}}{a^2}\delta_{ij} \label{eq:ADMinvmetric}
\end{align}
and
\begin{align}\label{eq:ADMconstraints}
N = 1 + \frac{\dot{\mathcal{R}}}{H}\, , && N^i = -\frac{\partial_i\mathcal{R}}{a H} + \epsilon\partial_i\nabla^{-2}\dot{\mathcal{R}}\, .
\end{align}
Here the derivative with respect to comoving coordinate $x^i$ is indicated as $\partial_i$. The relevance of the quantity $\mathcal{R}$ is that it represents the combination of adiabatic scalar metric perturbations that is time-independent for Fourier modes outside of the horizon, in the limit $k/aH \rightarrow 0$. Hence, it provides a convenient description of the initial conditions for perturbations from inflation.

In the interaction picture, the free field $\mathcal{R}$ can be expanded in annihilation and creation operators $\alpha(\mathbf{k})$ and $\alpha(\mathbf{k})^\dagger$ that satisfy $\left[\alpha(\mathbf{k}),\alpha^\dagger(\mathbf{k}') \right]= (2 \pi)^3 \delta^3(\mathbf{k}-\mathbf{k}')$:
\begin{equation}
\mathcal{R}(\mathbf{x}, \eta) = \int \frac{d^3 k}{(2 \pi)^3} \; \Bigl[ \mathcal{R}_k(\eta) e^{i\mathbf{k}\cdot\mathbf{x}} \alpha(\mathbf{k}) + \text{H.c.} \Bigr]
\end{equation}
and its time evolution is described by the Mukhanov-Sasaki equation~\cite{Mukhanov:1985rz,Sasaki:1986hm}:
\begin{equation}\label{eq:Mukhanov-Sasaki}
\mathcal{R}''_k + 2 aH(1+\delta+\epsilon)\mathcal{R}'_k+ k^2 \mathcal{R}_k = 0
\end{equation}
where $\delta =\ddot{H}/2H\dot{H}$ and $\epsilon=-\dot{H}/H^2$. In the exponential model for the inflaton potential under consideration here, $\delta=-\epsilon$ and Eq.~(\ref{eq:Mukhanov-Sasaki}) can be solved exactly in the case of constant $\epsilon$. The solution with the correct asymptotic behavior inside the horizon is:
\begin{eqnarray}
\mathcal{R}_k(\eta) &=& - \pi (1-\epsilon) e^{i\pi\nu/2+i\pi/4} \frac{H(\eta)}{\sqrt{\epsilon}M_\text{Pl}}\, 
\frac{(-k\eta)^{\frac{3}{2} }}{k^{3/2}} H^{(1)}_\nu(-k\eta)\\
\label{eq:Rcaltau}
&=& \mathcal{R}_{k*} u_{\nu}(-k \eta)
\end{eqnarray}
with $\nu = 3/2+\epsilon/(1-\epsilon)$, which satisfies the Wronskian condition $\mathcal{R}_k \mathcal{R}'^*_k-\mathcal{R}'_k \mathcal{R}^*_k = i (H/\varphi')^2$. The constant coefficient $ \mathcal{R}_{k*}$ is the value of the mode function outside the horizon,
$ \mathcal{R}_{k}(\eta) = \mathcal{R}_{k*}(\eta)$ for $-k\eta \ll 1$,
\begin{equation}
\mathcal{R}_{k*} = i e^{i\pi\nu/2+i\pi/4}(1-\epsilon)2^\nu \Gamma(\nu) 
\frac{H(\eta)}{ \sqrt{\epsilon} M_{\text{Pl}} } \frac{ (-k \eta)^{\frac{3}{2}-\nu} }{ k^{3/2} }\, .
\end{equation}
Since the $\eta$-dependence above cancels out, $H(\eta_I) (-k \eta_I)^{\frac{3}{2}-\nu} = H(\eta_*) (-k \eta_*)^{\frac{3}{2}-\nu}$, in which case the mode function for Fourier modes outside the horizon may be recast in terms of some other reference time or scale indicated by the asterisk:
\begin{equation}
\mathcal{R}_{k*} = i e^{i\pi\nu/2+i\pi/4}(1-\epsilon)^{\nu-\frac{1}{2}} \frac{2^\nu \Gamma(\nu) }{k^{3/2}}
\frac{H_*}{ \sqrt{\epsilon} M_{\text{Pl}} }  \left( \frac{k}{a_* H_*} \right)^{\frac{3}{2}-\nu}  \, .
\end{equation}
where we have also used $\eta_* = -1/(1-\epsilon) a_* H_*$. (A pedagogical discussion of this model, including the background evolution equations and the behavior of the gauge invariant $\mathcal{R}$ can be found in \cite{Weinberg:2008zzc}, sections 4.2 and 10.1, although our conventions are slightly different.) The two-point correlation function is used to define the power spectrum $P_{\cal R}$:
\begin{equation}
\langle \mathcal{R} (\mathbf{x},\eta) \mathcal{R} (\mathbf{y},\eta) \rangle =  \int \frac{d^3 k}{(2 \pi)^3} e^{i\vec k \cdot (\vec x- \vec y)} P_ \mathcal{R}(k), \qquad
P_ \mathcal{R}(k) =  | \mathcal{R}_{k}|^2.
\end{equation}
The power spectrum for modes outside the horizon is
\begin{equation}
\Delta^2_{\cal R}(k) \equiv \frac{k^3}{2\pi^2} P_ \mathcal{R}(k)  =  C_\nu  \frac{H^2_{*}}{\epsilon M_\text{Pl}^2} \left( \frac{k}{a_* H_*}\right)^{n_S-1}\, 
\end{equation}
where $C_\nu = 2^{2\nu-1}(1-\epsilon)^{2\nu-1}\Gamma(\nu)^2 /\pi^2$ and $n_S = 4 - 2\nu = 1 - 2 \epsilon/(1-\epsilon)$. The 7-year WMAP data \cite{Komatsu:2010fb} constrain the parameter values $n_S = 0.967\pm0.014$ and $C_\nu H_*^2/\epsilon M_\text{Pl}^2 = 2.43 ( \pm 0.091 ) \times 10^{-9}$ at an inverse-length scale $a_* H_* (= k_0) = 0.002 ~ {\rm Mpc}^{-1}$, so that $\epsilon = 0.0162 \pm 0.0067$ and $H_* = 1.12 (\pm 0.23) \times 10^{-5} ~M_\text{Pl}$.

%%%%%%%%%%%%%%%%%%%%%%%%%%%%%%%%%%%%%%%%%%%%%%%%%%%%%%%%%%%%%%
\section{\label{sec:threepointfunc}Correlation of magnetic fields and curvature perturbations}

We now evaluate the cross correlation between the primordial magnetic field and the scalar curvature perturbation, ${\cal R}$, making use of the in-in formalism \cite{Maldacena:2002vr, Weinberg:2005vy,Weinberg:2006ac}. The expectation value of a cosmological observable $\mathcal{O}$ in the state of the universe is given in perturbation theory by the standard rules of the in-in formalism:
\begin{equation}\label{eq:expectationOinin}
\langle \mathcal{O}_h(t) \rangle = \left\langle \left( T \exp\left( -i \int^t_{-\infty+} dt'\, H_{int}(t') \right) \right)^\dagger \mathcal{O}^{int}(t)\; T \exp\left( -i \int^t_{-\infty +} dt''\, H_{int}(t'') \right) \right\rangle\, ,
\end{equation}
where $\mathcal{O}_h$ is the Heisenberg picture operator, $H_{int}$ is the interaction Hamiltonian and $\mathcal{O}^{int}$ is the field in the interaction picture. (The time integrals are finite in perturbation theory after an appropriate Wick rotation or an analytic continuation into the complex plane that introduces an effective small positive imaginary part $i\epsilon'$ to time, which we indicate with the lower limit $\infty+$.) 

The interaction Hamiltonian is obtained by expanding the action to linear order in perturbations. We work in the so-called Maldacena gauge where $\delta\varphi=0$, which is convenient because perturbations of $W$ vanish. Then, expanding the interaction term in Eq.~(\ref{eq:actionWFF}) to linear order in $\mathcal{R}$,
\begin{eqnarray}
S_{int} &=& -\frac{1}{4}\int d^4 x \left[ (\sqrt{-g})^{(1)} \left( g^{\mu\lambda} g^{\nu\sigma} \right)^{(0)} 
+ (\sqrt{-g})^{(0)} \left(  g^{\mu\lambda} g^{\nu\sigma} \right)^{(1)}
\right] W(\varphi)   F_{\mu\nu} F_{\lambda\sigma} \\ \cr
&=&  \int d\eta\, d^3x \,W(\varphi) \left\{ \left( 3 \mathcal{R} + \frac{\mathcal{R}'}{a H}\right)  \left(\frac{1}{2} {A'_i}^2 - \frac{1}{4} F_{ij}^2 \right)\right. \\ \cr
&&\qquad -  \left.  \left[ \left(\mathcal{R} + \frac{\mathcal{R}'}{a H}\right){A_i'}^2 - \mathcal{R}F_{ij}^2 - \frac{\partial_i\mathcal{R}}{a H}A_j' F_{ij} + \epsilon \partial_i(\nabla^{-2}\mathcal{R}')A_j' F_{ij}\right]\right\}
\end{eqnarray}
where we have used Eqs.~(\ref{eq:ADMmetric})--(\ref{eq:ADMconstraints}). Putting these pieces together, and neglecting the term of order $\epsilon$, the interaction Hamiltonian in conformal time is given by
\begin{equation}\label{eq:Hint}
H_{int} = -\frac{1}{2} \int d^3 x \; W(\eta) \Biggl[\left(\mathcal{R} - \frac{\mathcal{R}'}{aH} \right)\left(A'^2_i + \frac{1}{2} F_{ij}^2\right)
 + \frac{2(\partial_i \mathcal{R})}{aH}A'_j F_{ij} \Biggr] .
\end{equation} 
This Hamiltonian differs from that used in the toy model of our earlier work. In particular, if we eliminate the $\mathcal{R}'$ and $\partial_i\mathcal{R}$ terms and substitute $\delta\phi$ for $\mathcal{R}$, then Eq.~(\ref{eq:Hint}) would resemble the earlier Hamiltonian. 

The cross correlation is
\begin{equation}
\langle \mathcal{R}(\mathbf{x},\eta) B_i(\mathbf{y},\eta) B^{i}(\mathbf{z}, \eta)\rangle = - 2 \,\text{Im}\; \int^{\eta}_{-\infty+} d\eta_1 \langle H_{int}(\eta_1)\mathcal{R}(\mathbf{x},\eta) B_k(\mathbf{y},\eta) B^{k}(\mathbf{z}, \eta)\rangle\, .
\end{equation}
In terms of Fourier-space three point function $P_3$,
\begin{equation}\label{eq:Gfourier}
\left\langle\mathcal{R}(\mathbf{x}, \eta) B_i(\mathbf{y},\eta) B^i(\mathbf{z}, \eta) \right\rangle = \int \left[ \prod^3_{i} \frac{d^3 k_i}{(2 \pi)^3} \right] e^{-i \mathbf{k}_1 \cdot\mathbf{x}- i \mathbf{k}_2 \cdot\mathbf{y} - i \mathbf{k}_3 \cdot\mathbf{z}} \, (2\pi)^3 \delta^3(\mathbf{k}_1+\mathbf{k}_2+\mathbf{k}_3) P_3(k_1,k_2,k_3; \eta)\, ,
\end{equation}
with
\begin{equation}\label{eq:Gstep1}
P_3(k_1,k_2,k_3; \eta) = - 2 \, \text{Im} \left\{ \frac{\mathcal{R}_{k_1*}v_{k_2*} v_{k_3*}}{ a(\eta)^{4} } \int^\eta_{-\infty+} d\eta_1 \, \mathcal{I} \right\}
\end{equation}
where the external lines are taken outside the horizon. Then, from the Hamiltonian, the structure of $\mathcal{I}$ is found to be
\begin{eqnarray}\label{eq:Ical}
\mathcal{I} &=& W(\eta_1) \left[ \left(\mathcal{R}_{k_1}(\eta_1) - \frac{\mathcal{R}'_{k_1}(\eta_1)}{a(\eta_1) H(\eta_1)}\right)  
 \left(2 k_2 k_3\mu  v'_{k_2}(\eta_1) v'_{k_3}(\eta_1) - k_2^2 k_3^2 (1+\mu^2)v_{k_2}(\eta_1) v_{k_3}(\eta_1)\right) \right. \cr
 &&\qquad\qquad + \left. \frac{k_2^2 k_3^2}{a(\eta_1)H(\eta_1)} \mathcal{R}_{k_1}(\eta_1) \left(  \vartheta(k_2,k_3,\mu)v_{k_2}(\eta_1) v'_{k_3}(\eta_1) + \vartheta(k_3,k_2,\mu)v'_{k_2}(\eta_1) v_{k_3}(\eta_1)  \right) \right] 
\end{eqnarray}
where we have introduced $\mu = \cos\theta = \hat k_2 \cdot \hat k_3$ and $\vartheta(x,y,\mu)=1 +\mu^2 +2\mu (x/y)$. Details of this calculation can be found in the Appendix. With the aid of Eqs.~(\ref{eq:Rcaltau}) \& (\ref{eq:vtau}), one can see that $P_3$ will contain a factor $\vert \mathcal{R}_{k_1*}\vert^2 \vert v_{k_2*} \vert^2 \vert v_{k_3*}\vert^2$ and a sum of the real parts of time integrals of products of Hankel functions and their derivatives. Specifically,
\begin{eqnarray}\label{eqn:P3general}
P_3(k_1,k_2,k_3; \eta) &=&   \vert \mathcal{R}_{k_1*}\vert^2 \vert v_{k_2*}\vert^2 \vert v_{k_3*}\vert^2  \frac{W_I}{a^4(\eta)}   \cr
&& \times
\tau_I^{2n/(1-\epsilon)} \left(K_1 \mu + K_2 (1 + \mu^2) + K_{3a} \vartheta(x_2,x_3,\mu)+ K_{3b} \vartheta(x_3,x_2,\mu) \right) \\
\label{eqn:K1general}
K_1 &=& -4 k_1 k_2 k_3 \,
\, \text{Im}\left\{ 
\int_{\tau}^\infty d\tau_1  \tau_1^{-2n/(1-\epsilon)} \,
\frac{d}{d\tau_1} u_\alpha(x_2 \tau_1) \, \frac{d}{d\tau_1} u_\alpha(x_3 \tau_1) \, \left( 1 +(1-\epsilon)\tau_1 \frac{d}{d\tau_1}\right)u_\nu(\tau_1)  \right\} \\
\label{eqn:K2general}
K_2 &=& 2 \frac{k_2^2 k_3^2}{k_1} \,
\, \text{Im}\left\{ \int_{\tau}^\infty d\tau_1  \tau_1^{-2n/(1-\epsilon)} \,
u_\alpha(x_2 \tau_1) u_\alpha(x_3 \tau_1) \, \left( 1 +(1-\epsilon)\tau_1 \frac{d}{d\tau_1}\right)u_\nu(\tau_1) \right\}  \\
\label{eqn:K3general}
K_{3a} &=& 2 \frac{k_2^2 k_3^2}{k_1} \,
\, \text{Im}\left\{ \int_{\tau}^\infty d\tau_1  \tau_1^{-2n/(1-\epsilon)} \,
u_\alpha(x_2 \tau_1) (1-\epsilon) \tau_1 \frac{d}{d\tau_1}u_\alpha(x_3 \tau_1)  u_\nu(\tau_1)  \right\}
\end{eqnarray}
where $x_2=k_2/k_1,\, x_3=k_3/k_1$, $\tau = - k_1\eta$ and $K_{3b} = K_{3a}(k_2\leftrightarrow k_3)$.

For general slow-roll parameter $\epsilon$, these integrals cannot be evaluated in an analytical closed form, and require numerical evaluation. However, the calculation is transparent for the $n=2$ case neglecting the slow-roll parameter $\epsilon$ in the order of the Hankel functions, i.e. we take $\nu=3/2$ and $\alpha =5/2$ inside the mode functions. In the appendix we list all integrals and their explicit form in this case. 

The amplitude $P_3$ at the end of inflation, in the case $n_B=0$ and $\epsilon \ll 1$, is
\begin{eqnarray}\label{eq:Gfinal}
P_3(k_1, k_2, k_3,\eta_I) &=&
\vert \mathcal{R}_{k_1*}\vert^2 \vert v_{k_2*}\vert^2 \vert v_{k_3*}\vert^2  W_I  \tau_I^4 /a^4(\eta_I)\cr\cr
&& \times \left( \mu K_1 + (1 + \mu^2)K_2 +  \vartheta(x_2,x_3,\mu) K_{3a}
+  \vartheta(x_3,x_2,\mu) K_{3b} \right) \\ \cr
K_1 &=& -\frac{4}{9} \frac{k_2^3 k_3^3}{k_1^4 \omega^3}\left[  
-k_2 k_3 ( \omega^2 + k_1 \omega -2 k_1^2 ) +\omega(\omega^3 - k_1 \omega^2 + 3 k_1^2\omega-k_1^3)
  \right]  \\
K_2 &=& 2  \frac{k_2^2 k_3^2}{k_1} \left[ -\frac{4}{3} \left(\gamma + \ln[-\omega\eta_I] \right) + \frac{1}{9 k_1^3 \omega^3}\left(
k_2^2 k_3^2 (\omega^2 +k_1 \omega - 2 k_1^2) \right. \right. \cr
&& \left.  \left.- 3 k_2 k_3 \omega(\omega^3 - k_1 \omega^2 + 3k_1^2\omega - k_1^3)
-3 k_1 \omega^2(\omega^3 - 4 k_1 \omega^2 -2 k_1^2 \omega + k_1^3) \right) \right]  \\
K_{3a} &=& \frac{2}{9} \frac{k_2^2 k_3^4}{k_1^4 \omega^3}  \left[ 3 \omega^4 - 3\omega^3(k_1+k_2) +\omega^2(3 k_1^2 + k_2^2) +\omega k_2(3 k_1^2  + k_1 k_2 + k_2^2) + 2 k_1 k_2^2(k_1+k_2)\right]   
\end{eqnarray}
and $K_{3b} = K_{3a}(x_2\leftrightarrow x_3)$,
where $\omega= k_1+k_2+k_3$ and $\gamma$ is the Euler-Mascheroni constant.

We note that, although the individual Fourier modes contributing to the three-point function are constant outside the horizon,
the correlation continues to grow logarithmically. In our case, the contribution denoted $K_2$ introduces such a correction surprisingly \textit{at tree level}. The term $K_2$ is the analogue to the $\log(k/aH)$ contribution found in our previous work for the cross correlation of a spectator scalar field with the magnetic field energy density --- see $\mathcal{I}_2$ for $n=2$ studied in \cite{Caldwell:2011ra}. Both of these results are rather puzzling since one would expect $\log$ corrections to appear only from loops. 

It is appropriate to pause here and note that Eqs.~(\ref{eqn:P3general}-\ref{eqn:K3general}) for the amplitude $P_3$ will be valid even in the absence of an amplification mechanism, {\it i.e.} even when $W(\eta)=1$.  The result in that case can be obtained by simply setting $n=0$ and $W_I=1$.  Carrying out this calculation, we find
\begin{eqnarray}
K_1 &=& 2 K_2 = 4\frac{k_2^2 k_3^2}{\omega^3}(\omega-k_1)(\omega+2 k_1)\times(1 + {\cal O}(\tau)) \\ \cr
K_{3a} &=& -2\frac{k_2^2 k_3^3}{\omega^3}(\omega+2 k_1)\times(1 + {\cal O}(\tau)) ,\, \qquad K_{3b} = K_{3a}(k_2 \leftrightarrow k_3).
\end{eqnarray}
However, the sum yields $P_3=0$ to the order of approximation of our calculation. Since $\tau \ll 1$ for modes outside the horizon at the end of inflation, any non-zero cross-correlation is exceedingly small in the absence of an amplification mechanism.

%%%%%%%%%%%%%%%%%%%%%%%%%%%%%%%%%%%%%%%%%%%%%%%%%%%%%%%%%%%%%%
\section{\label{sec:bispectrum}Behavior of the cross-correlation}

We proceed to analyze the cross-correlation between the primordial magnetic field and the curvature perturbation, to determine if there is any imprint of unique signature to indicate the amplification mechanism.

%%%%%%%%%%%%%%%%%%%%%%%%%%%%%%%%%%%%%%%%%%%%%%%%%%%%%%%%%%%%%%
\subsection{Real-Space Cross-Correlation Coefficient}

The cross-correlation amplitude, evaluated in the coincidence limit, can be determined as follows.  Starting from Eq.~(\ref{eq:Gfinal}), we evaluate the $\mathbf{k}_1$ integral to eliminate the delta function. The remaining integrand depends only on the magnitudes $k_2$, $k_3$, and $\theta$, the angle between the two vectors:
\begin{equation}
     \langle \mathcal{R} B^2 \rangle = \frac{1}{8 \pi^4} \int k_2^2 dk_2 \, k_3^2 dk_3 \, d(\cos\theta) \, P_3(k_1, k_2, k_3)
     \label{eqn:p3integral}
\end{equation}
where $k_1 = (k_2^2 + k_3^2 + 2 k_2 k_3 \cos\theta)^{1/2}$. However, we can replace the $\theta$ integral by $k_1$, whereby
\begin{equation}
      \langle \mathcal{R} B^2 \rangle = \frac{1}{8 \pi^4} \int
     k_2 dk_2 \, k_3 dk_3 \int_{|k_2 - k_3|}^{k_2 + k_3} k_1
     dk_1 \, P_3(k_1, k_2, k_3).
\end{equation}
Since the integrand is invariant under the exchange of $k_2$ and $k_3$, we can replace $P_3 \to 2 P_3 \theta(k_2 - k_3)$ and
remove the absolute-value sign from the lower limit of integration. We implement cutoffs at both large and small $k$, for the ultraviolet and infrared divergences that arise in both the curvature perturbation and magnetic field spectra. The cross correlation evaluates to
\begin{equation}
\langle \mathcal{R} B^2\rangle = \frac{36}{\pi^3} \left( \frac{H_I}{\sqrt{\epsilon} M_\text{Pl}}\right)^2 \frac{H_I^4}{W_I} (\log r)^2 ( N_I    - \frac{2}{3}\log r)
\end{equation}
where $N_I$ is the number of e-foldings of inflation, $r = k_\text{max}/k_\text{min}$,  and $k_\text{max}$ and $k_\text{min}$ are upper and lower bounds on the run of wavevectors. In practice, we expect to link the minimum wavevector with the Hubble scale, $k_\text{min} \simeq 2 \pi H_0$, and the maximum wavevector with some galactic scale, $k_\text{max} \simeq 2 \pi/\lambda$ where $\lambda\sim$kpc. Since $|k \eta_I |\ll 1$, we have discarded subdominant terms from the above results. The dimensionless cross-correlation coefficient $X_{ \mathcal{R} B^2}$, formed from the ratio of the cross correlation with the root-mean-square amplitudes of the curvature perturbation and magnetic fields gives
\begin{equation}
X_{\mathcal{R} B^2} = \frac{\langle \mathcal{R} B^2\rangle}{\langle \mathcal{R}^2\rangle^{1/2} \langle B^2\rangle}
= \frac{8}{\sqrt{\pi}} \left( \frac{H_I}{\sqrt{\epsilon}  M_\text{Pl}} \right) \sqrt{\log r} (N_I - \frac{2}{3}\log r)
\end{equation}
Considering a sufficiently wide range of scales, e.g. $r \gtrsim 10^4$, and using $N_I \simeq 70$   then $X_{\mathcal{R} B^2} \simeq  900 \times ({H_I}/{\sqrt{\epsilon}  M_\text{Pl}})$, which is nearly three orders of magnitude larger than a naive expectation for the amplitude.

%%%%%%%%%%%%%%%%%%%%%%%%%%%%%%%%%%%%%%%%%%%%%%%%%%%%%%%%%%%%%%
\subsection{Discretized Fourier-Space Cross-Correlation Coefficient}

We now evaluate the triangle-shape dependence of the full three-point correlation function in discrete Fourier space.  As discussed in Sec. IV C of~\cite{Caldwell:2011ra}, this leads to a visualization of the cross correlation that has a clearer imprint of the amplification mechanism (cf. Fig.~1 of \textit{ibid}.). We evaluate a ratio of the form,
\begin{equation}
     \frac{P_3(k_1,k_2,k_3)}{\sqrt{P_{\mathcal{R}}(k_1) P_B(k_2) P_B(k_3)}},
\end{equation}
to normalize the cross-correlation power spectrum. Since this ratio is not dimensionless, however, we convert the continuous Fourier transforms into discretized Fourier transforms,
\begin{equation}
     \int \frac{d^3k}{(2 \pi)^3} \to \frac{1}{V}\sum_{\vec n},
\end{equation}
and likewise replace the Dirac delta function with a Kronecker delta,
\begin{equation}
     (2 \pi)^3 \delta(\vec k_1+ \vec k_2) \to V \delta_{\vec n_1,\vec n_2}.
\end{equation}
We presume a maximum length, $L$, so that the volume is $V=L^3$ and mode numbers are $k_i = 2 \pi n_i/L$. The curvature-perturbation and magnetic-field power spectra are now
\begin{eqnarray}
     \langle{ \mathcal{R}^2 }\rangle &=& \sum_{\vec n} e^{i\vec n \cdot (\vec x-\vec y)/L } \widetilde P_{\mathcal{R}},\\ 
     \widetilde P_{\mathcal{R}} &=& V^{-1} P_{\mathcal{R}}, \\
     \langle{ B^2}\rangle &=& \sum_{\vec n} e^{i\vec n \cdot( \vec x - \vec y)/L } \delta_{\vec n_1,\vec n_2} \widetilde P_{B},\\
     \widetilde P_{B} &=& V^{-1} P_{B},
\end{eqnarray}
so that $\widetilde P_{\mathcal{R}}$ is dimensionless and $\widetilde P_{B}$ has units of (energy)${}^4$. The three-point function becomes
\begin{eqnarray}
     \langle{\mathcal{R} B^2}\rangle &=& \sum_{\vec n_1+\vec n_2+\vec n_3=0} 
     e^{i(\vec n_1 \cdot \vec x + \vec n_2\cdot\vec y+ \vec n_3\cdot\vec z)/L }  \widetilde P_{3},\nonumber\\ 
     \widetilde P_{3} &=& V^{-2} P_{3},
\end{eqnarray}
where $\widetilde P_3$ has units of (energy)${}^4$. We can now build a dimensionless cross-correlation coefficient,
\begin{equation}
     C = \frac{\widetilde P_{3}(n_1,n_2,n_3)}{\sqrt{\widetilde P_{\mathcal{R}}(n_1) \widetilde P_B(n_2) \widetilde P_B(n_3)}},
\end{equation}
where $n_i$ for $i=1,2,3$ are the magnitudes of vectors $\vec n_i$ that form a closed triangle.

\begin{figure*}[htbp]
\vspace{20pt}
\begin{center}
\includegraphics[scale=1.0]{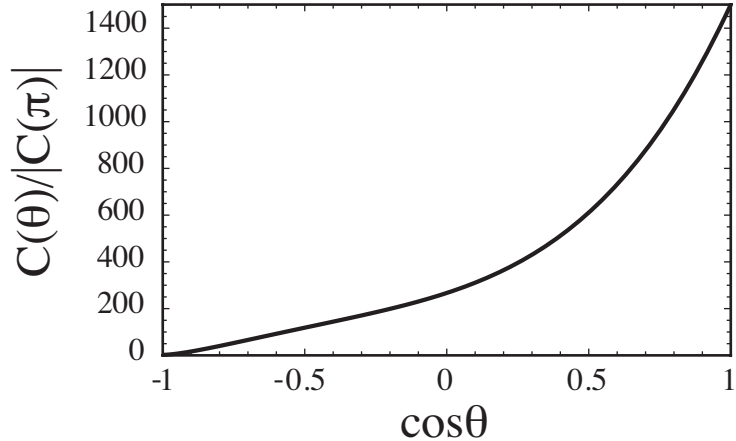}
\end{center}
\caption{The ratio $C(\theta)/|C(\pi)|$ is shown as function of $\cos\theta$. We have set $\delta = 2\pi n_1 |\eta_I/L| \sim 10^{-27}$
for horizon-sized modes. The flattened triangle, at $\cos\theta=1$, yields a cross-correlation amplitude $C(\theta=0)\simeq 1500\, |C(\pi)|$.}
\label{fig:figure1}
\end{figure*}

For isosceles triangles with $n_2 = n_3$, the correlation $C$ has two interesting limits. 
First, for a squeezed triangle, with $1 \le n_1 \ll n_2$ or $\theta = \pi$,  then
\begin{equation}
C(\theta=\pi) = -\frac{2}{\pi (n_1)^{3/2}}\left( \frac{H_I}{\sqrt{\epsilon} M_\text{Pl}}\right).
\end{equation}
Hence, the curvature perturbation and magnetic fields are anti-correlated for squeezed triangles.
Second, for a flattened triangle with $n_2=n_1/2$ or $\theta=0$, then
\begin{equation}
C(\theta= 0) = -\frac{2}{\pi (n_1)^{3/2}}\left( \frac{H_I}{\sqrt{\epsilon} M_\text{Pl}}\right) 
\times 24 \left( \ln(2 \delta n_1) + \gamma - \frac{3}{2}\right)
\end{equation}
where $\delta = -2 \pi \eta_I/L \ll 1$. Since $\delta n_1 \sim 10^{-27}$ for horizon-sized modes, then it is clear that the cross correlation for flattened triangle configuration is not only positive but much greater than for the squeezed triangle. A plot showing the angular dependence of the cross correlation is shown in Fig.~\ref{fig:figure1}.
 
\begin{figure*}[htbp]
\vspace{20pt}
\begin{center}
\includegraphics[scale=1.0]{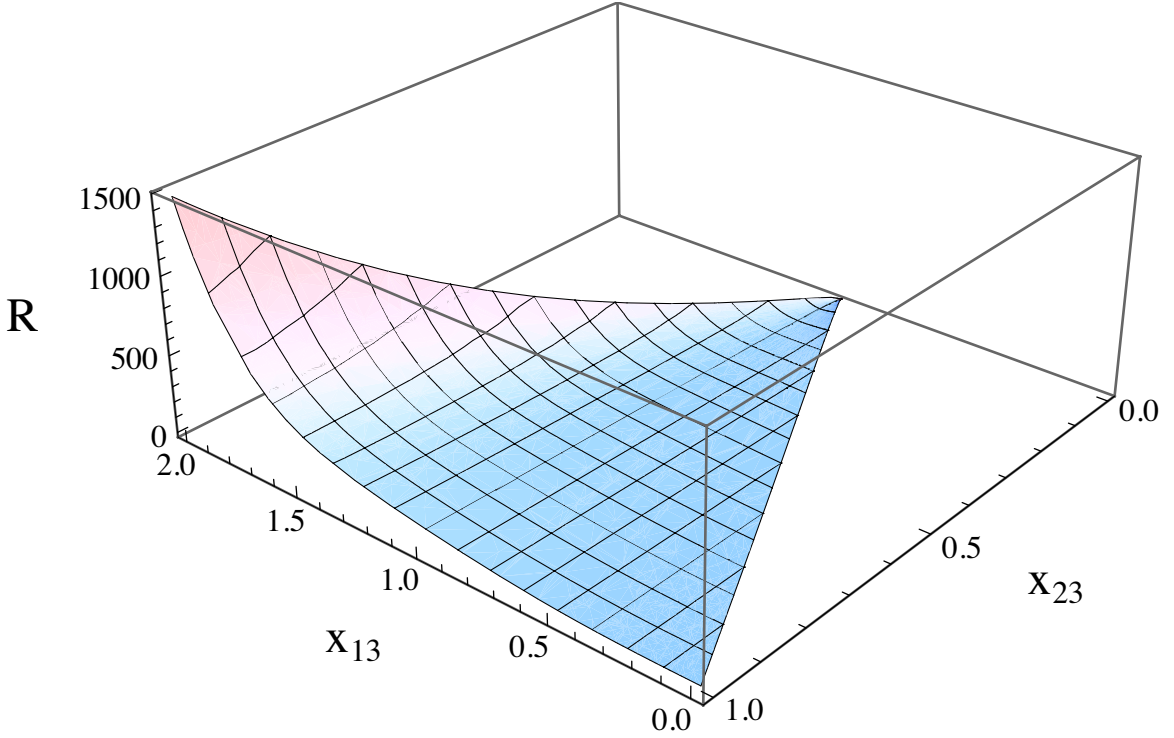}
\end{center}
\caption{The quantity $R$, defined in the text as the ratio of the Fourier-space cross-correlation coefficient to that of the squeezed triangle, times a factor $x_{23}^2$, is shown as a function of the triangle side lengths.  We have set $\delta = 2\pi n_1 |\eta_I/L| \sim 10^{-27}$ for horizon-sized modes. The flattened triangle, at $\cos\theta=1$, yields a cross-correlation amplitude $C(\theta=0)\simeq 1500\, |C(\pi)|$.}
\label{fig:figure2}
\end{figure*}

To show the full Fourier-space triangle dependence of the cross-correlation, we define the quantity
\begin{equation}
R \equiv  \left(\frac{n_2}{n_3}\right)^{2} \frac{C(\theta)}{|C(\pi)|}
\end{equation}
and introduce the variables $x_{23} \equiv n_2/n_3$ and $x_{13} \equiv n_1/n_3$, where $0\le x_{23} \le 1$ and $1-x_{23} \le
x_{13} \le 1+x_{23}$ covers the full set of triangles.  As seen in Fig.~\ref{fig:figure2}, the maximum value of $R$ occurs for the flattened triangles, corresponding to the line $x_{13} = 1+ x_{23}$, along which $\theta=0$. Squeezed triangles, where $\theta = \pi$, are located along $x_{13} = 1-x_{23}$.

%%%%%%%%%%%%%%%%%%%%%%%%%%%%%%%%%%%%%%%%%%%%%%%%%%%%%%%%%%%%%%
\section{\label{sec:conclusions}Conclusions and Outlook}

Understanding the origin of galactic magnetism is one key aspect of the evolution and structure of galaxies, and has attracted much interest. Single field slow-roll inflation can in general accommodate a natural extension of electrodynamics that leaves a random cosmic magnetic field outside the horizon at the end of inflation that is large enough to seed the galactic dynamo. In this work we studied the dominant primordial non-gaussian contribution to the statistics of this field. This comes from the cross correlation of metric scalar perturbations and magnetic field energy density, $\langle \mathcal{R}B^2\rangle$. We found that in power-law inflation when electrodynamics has a dilaton coupling of the inflaton background, there exists a spectrum $\Delta_{\mathcal{R}BB}$ of order $10^{-5}$, which is found to peak for flattened triangles in Fourier space.

We expect two classes of experiments may be sensitive to this primordial non-gaussianity: deep-field all-sky Faraday rotation measures and polarization detection of the CMB. For the first class of experiments, examples are SKA~\cite{Beck:2009ew,ars-5-399-2007,Gaensler:2006tj} and LOFAR~\cite{Beck:2009ew,Rottgering:2006ms} which have projected sensitives close to $10^{-10}\,\text{G}$.  In the SKA experiment, the relevant effect arises from the fact that $\mathcal{R}$ is directly proportional to the dark matter contrast, hence $\langle\mathcal{R}B^2\rangle$ constitutes the primordial cross correlation of matter density and cosmic magnetic fields. Preliminary searches and studies of this correlation include the Refs.~\cite{Stasyszyn:2010kp,Xu:2005rb}.

In the CMB, cosmic magnetic fields affect most prominently the polarization through the Faraday effect. The detectability and bounds on cosmic magnetic field power spectrum in the CMB have been extensively studied in the literature: a partial list of recent work relevant to Mpc wavelength fields is Refs.~\cite{Mack:2001gc,Lewis:2004ef,Shaw:2010ea,Kahniashvili:2010wm,Paoletti:2010rx,Kunze:2010ys}. Noting that the temperature fluctuation of the CMB is largely determined by $\mathcal{R}$ whereas the B-fields produce a frequency-dependent rotation of the polarization along the line of sight, then the cross correlation $\langle\mathcal{R}B^2\rangle$ introduces a primordial non-gaussianity in the temperature-polarization-polarization correlation. The feasibility of detecting such a signal remains to be studied.

%%%%%%%%%%%%%%%%%%%%%%%%%%%%%%%%%%%%%%%%%%%%%%%%%%%%%%%%%%%%%%
\appendix

%%%%%%%%%%%%%%%%%%%%%%%%%%%%%%%%%%%%%%%%%%%%%%%%%%%%%%%%%%%%%%
\section{Calculation of  ${\cal I}$}

We simplify our calculation by constructing the following differential operators. First, to obtain the magnetic field correlation from the vector potential correlation, we introduce 
\begin{equation}
{\cal O}_{B}^{lm} = \frac{1}{a^4}
\left( \delta^{lm} \frac{\partial^2}{\partial y^k \partial z^k} - \frac{\partial^2}{\partial y^m \partial z^l} \right)
\end{equation}
such that $\vec B(\vec y)\cdot \vec B(\vec z) = {\cal O}_{B}^{lm} A_{l}(\vec y) A_{m}(\vec z)$. Second, to obtain the interaction Hamiltonian density,
\begin{eqnarray}
{\cal O}_{(H1)}^{ij} &=& -\frac{1}{2}(1 - \frac{1}{a(\eta_1) H(\eta_1)} \frac{\partial}{\partial \eta_1}) \frac{\partial}{\partial\eta_2}\frac{\partial}{\partial \eta_3} \delta^{ij} \\\cr
{\cal O}_{(H2)}^{ij} &=& -\frac{1}{2}(1 - \frac{1}{a(\eta_1) H(\eta_1)} \frac{\partial}{\partial \eta_1}) \frac{\partial}{\partial x_2^k }\frac{\partial}{\partial x_3^l}
\left( \delta^{ij} \delta^{kl} - \delta^{il} \delta^{jk}\right)\\ \cr
{\cal O}_{(H3)}^{ij} &=& -\frac{1}{2 a(\eta_1) H(\eta_1)} \frac{\partial}{\partial x_1^k} \left[ \left(\delta^{kl} \delta^{ij} - \delta^{il} \delta^{jk}\right) \frac{\partial}{\partial \eta_2}\frac{\partial}{\partial x_3^l} + \left(\delta^{kl} \delta^{ij} - \delta^{ik} \delta^{jl}\right) \frac{\partial}{\partial \eta_3}\frac{\partial}{\partial x_2^l} \right]
\end{eqnarray}
such that 
\begin{equation}
{\cal H}_{int} = \lim_{\substack{x_{2,3}\to x_1 \\ \eta_{2,3}\to \eta_1}} 
\left({\cal O}_{(H1)}^{ij}+{\cal O}_{(H2)}^{ij}+{\cal O}_{(H3)}^{ij}\right) 
{\cal R}(\mathbf{x}_1,\eta_1)A_i(\mathbf{x}_2,\eta_2) A_j(\mathbf{x}_3,\eta_3) 
\end{equation}
where $H_{int} = \int d^3 x_1 a^4(\eta_1) {\cal H}_{int}$.
Using these definitions, we can rewrite the cross correlation as
\begin{eqnarray}
\langle \mathcal{R}(\mathbf{x},\eta) B_i(\mathbf{y},\eta) B^{i}(\mathbf{z}, \eta)\rangle 
&=&  - 2 \,\text{Im}\; \int^{\eta}_{-\infty+} d\eta_1 \langle H_{int}(\eta_1)\mathcal{R}(\mathbf{x},\eta) B_k(\mathbf{y},\eta) B^{k}(\mathbf{z}, \eta)\rangle \cr \cr
&=& -2 \,\text{Im}\; \int^{\eta}_{-\infty+} d\eta_1 \int d^3 x_1 W(\eta_1) 
 \lim_{x_{2,3}\to x_1}\left({\cal O}_{(H1)}^{ij}+{\cal O}_{(H2)}^{ij}+{\cal O}_{(H3)}^{ij}\right) {\cal O}_{B}^{lm} \cr \cr
 &&\qquad \times
\langle {\cal R}(\mathbf{x}_1,\eta_1) A_i(\mathbf{x}_2,\eta_2) A_j(\mathbf{x}_3,\eta_3) 
 {\cal R}(\mathbf{x},\eta) A_l(\mathbf{y},\eta) A_m(\mathbf{z},\eta) \rangle \, .
\end{eqnarray}
The expectation value in the above expression breaks into $\langle {\cal R}(\mathbf{x}_1,\eta_1){\cal R}(\mathbf{x},\eta)\rangle$
and $\langle A_i A_j A_l A_m \rangle$, where
\begin{equation}
\langle {\cal R}(\mathbf{x}_1,\eta_1) {\cal R}(\mathbf{x},\eta)\rangle = \int \frac{d^3 k_1}{(2 \pi)^3} {\cal R}_{k_1}(\eta_1) {\cal R}_{k_1}^{*}(\eta) e^{i \mathbf{k}_1 \cdot(\mathbf{x}_1 - \mathbf{x})}
\end{equation}
and
\begin{eqnarray}
\langle A_i(\mathbf{x}_2,\eta_2) A_j(\mathbf{x}_3,\eta_3) A_l(\mathbf{y},\eta) A_m(\mathbf{z},\eta)\rangle &=& \int \frac{d^3 k_2}{(2 \pi)^3}\frac{d^3 k_3}{(2 \pi)^3} v_{k_2}^{*}(\eta)  v_{k_3}^{*}(\eta)  \cr\cr
&&\times \left[ v_{k_2}(\eta_2)  v_{k_3}(\eta_3)  e^{i \mathbf{k}_2 \cdot(\mathbf{x}_2 - \mathbf{y})+i \mathbf{k}_3 \cdot(\mathbf{x}_3 - \mathbf{z})} P_{il}(\hat k_2)  P_{jm}(\hat k_3)  \right. \cr\cr
&&+ 
\left. v_{k_2}(\eta_3)  v_{k_3}(\eta_2)  e^{i \mathbf{k}_2 \cdot(\mathbf{x}_3 -\mathbf{y})+i \mathbf{k}_3 \cdot(\mathbf{x}_2 - \mathbf{z})} P_{jl}(\hat k_2) P_{im}(\hat k_3) \right]
\end{eqnarray}
with $P_{ij}(\hat k) = \delta_{ij} - (\hat k)_i (\hat k)_j$. In the above expressions and below, a superscript asterisk indicates the complex conjugate of a number, not to be confused with a subscript indicating a quantity outside the horizon.

We can now begin to evaluate the cross correlation. The operator ${\cal O}_B$ acting on the vector potential four-point function gives 
\begin{eqnarray}
{\cal O}_{B}^{lm} \langle A_i(\mathbf{x}_2,\eta_2) A_j(\mathbf{x}_3,\eta_3) A_l(\mathbf{y},\eta) A_m(\mathbf{z},\eta)\rangle &=&
\frac{1}{a^4(\eta)}\int \frac{d^3 k_2}{(2 \pi)^3}\frac{d^3 k_3}{(2 \pi)^3} v_{k_2}^{*}(\eta)  v_{k_3}^{*}(\eta)  k_2 k_3 \cr\cr
&& \times \left[ v_{k_2}(\eta_2)  v_{k_3}(\eta_3)  e^{i\mathbf{k}_2 \cdot(\mathbf{x}_2 - \mathbf{y})+i\mathbf{k}_3 \cdot(\mathbf{x}_3 - \mathbf{z})}  
( (\hat k_3)_i  (\hat k_2)_j  - \mu \delta_{ij} )  \right. \cr\cr
&& +  \left. v_{k_2}(\eta_3)  v_{k_3}(\eta_2)  e^{i \mathbf{k}_2 \cdot(\mathbf{x}_3 - \mathbf{y})+i \mathbf{k}_3 \cdot(\mathbf{x}_2 - \mathbf{z})}  
( (\hat k_2)_i  (\hat k_3)_j  - \mu \delta_{ij} )  \right]
\end{eqnarray}
where $\mu = \hat k_2 \cdot \hat k_3 = \cos\theta$. Next, including the coupling function $W$ and applying the ${\cal O}_{H}$ operators, 
\begin{eqnarray}
&& \lim_{\substack{x_{2,3}\to x_1 \\ \eta_{2,3}\to \eta_1}}  \, \sum_{n=1}^3 {\cal O}_{(Hn)}^{ij} {\cal O}_{B}^{lm} W(\eta_1)
\langle {\cal R}(\mathbf{x}_1,\eta_1) {\cal R}(\mathbf{x},\eta)\rangle \, \langle A_i(\mathbf{x}_2,\eta_2) A_j(\mathbf{x}_3,\eta_3) A_l(\mathbf{y},\eta) A_m(\mathbf{z},\eta)\rangle \cr\cr
&&\qquad =  \int \frac{d^3 k_1}{(2 \pi)^3}  \frac{d^3 k_2}{(2 \pi)^3}  \frac{d^3 k_3}{(2 \pi)^3} 
e^{i \mathbf{k}_1(\mathbf{x}_1-\mathbf{x})+i \mathbf{k}_2\cdot(\mathbf{x}_1 - \mathbf{y}) + i\mathbf{k}_3 \cdot(\mathbf{x}_1-\mathbf{z})} 
{\cal R}_{k_1}^{*}(\eta) v_{k_2}^{*}(\eta) v_{k_3}^{*}(\eta) \, {\cal I}
 \end{eqnarray}
where
\begin{eqnarray}
{\cal I} &=& W(\eta_1)\left\{ 2 \mu k_2 k_3 \left[ R_{k_1}(\eta_1) - \frac{R_{k_1}'(\eta_1)}{a_1 H_1}\right] v_{k_2}'(\eta_1) v_{k_3}'(\eta_1)\right. \cr \cr
&-&(1 + \mu^2) k_2^2 k_3^2 \left[ R_{k_1}(\eta_1) - \frac{R_{k_1}'(\eta_1)}{a_1 H_1}\right] v_{k_2}(\eta_1) v_{k_3}(\eta_1) \cr \cr
&+& \left. \frac{1}{a_1 H_1}k_2^2 k_3^2 R_{k_1}(\eta_1) 
\left[ v_{k_2}'(\eta_1)v_{k_3}(\eta_1) {\vartheta}(k_3,k_2,\mu) + v_{k_2}(\eta_1)v_{k_3}'(\eta_1) {\vartheta}(k_2,k_3,\mu)  \right]\right\}
\end{eqnarray}
and we define ${\vartheta}(x,y,\mu)= 1+ \mu^2 + 2\mu(x/y)$.  The cross correlation is then given by
\begin{eqnarray}
\langle \mathcal{R}(\mathbf{x},\eta) B_i(\mathbf{y},\eta) B^{i}(\mathbf{z}, \eta)\rangle 
&=& 
 - 2 \,\text{Im}\; \int^{\eta}_{-\infty+} d\eta_1 \int d^3 x_1 \frac{1}{a^4(\eta)}
 \int \frac{d^3 k_1}{(2 \pi)^3}  \frac{d^3 k_2}{(2 \pi)^3}  \frac{d^3 k_e}{(2 \pi)^3} \cr\cr
&& \times e^{i\mathbf{k}_1(\mathbf{x}_1-\mathbf{x})+i\mathbf{k}_2\cdot(\mathbf{x}_1 - \mathbf{y}) + i \mathbf{k}_3 \cdot(\mathbf{x}_1-\mathbf{z})} 
{\cal R}_{k_1}^{*}(\eta) v_{k_2}^{*}(\eta) v_{k_3}^{*}(\eta) \,  {\cal I}.
\end{eqnarray}
Rearranging the order of integration, and evaluating the $x_1$ integral, we may further simplify this expression to
\begin{eqnarray}
\langle \mathcal{R}(\mathbf{x},\eta) B_i(\mathbf{y},\eta) B^{i}(\mathbf{z}, \eta)\rangle 
&=&  \int \frac{d^3 k_1}{(2 \pi)^3}  \frac{d^3 k_2}{(2 \pi)^3}  \frac{d^3 k_3}{(2 \pi)^3}
e^{-i \mathbf{k}_1\cdot \mathbf{x} - i \mathbf{k}_2 \cdot \mathbf{y} - i \mathbf{k}_3 \cdot \mathbf{z} }  (2 \pi)^3 \delta^3(\mathbf{k}_1 + \mathbf{k}_2 + \mathbf{k}_3) P_3(k_1,k_2,k_3;\eta) \\
P_3(k_1,k_2,k_3;\eta) &=& - 2 \,\text{Im}\; \left\{ \frac{{\cal R}_{k_1}^{*}(\eta) v_{k_2}^{*}(\eta) v_{k_3}^{*}(\eta) }{a^4(\eta)}
\int^{\eta}_{-\infty+} d\eta_1  {\cal I} \right\}.
\end{eqnarray}
Hence we arrive at Eq.~(\ref{eq:Gstep1}).

%%%%%%%%%%%%%%%%%%%%%%%%%%%%%%%%%%%%%%%%%%%%%%%%%%%%%%%%%%%%%%
\section{Integrals}

The integrals required to evaluate the terms $K_i$ in Eqs.~(\ref{eqn:K1general}-\ref{eqn:K3general}) are given as follows:
\begin{eqnarray}
{J_{1a}}&=& {\rm Im}\,\left\{ \int_{\tau}^\infty d\tau_1 {\tau_1}^{-4} \,
\frac{d}{d\tau_1} u_{\frac{5}{2}}(x_2 \tau_1) \, \frac{d}{d\tau_1} u_{\frac{5}{2}}(x_3 \tau_1) \, u_{\frac{3}{2}}(\tau_1) \right\} \cr\cr
&=& \frac{ k_2^2 k_3^2}{9 k_1^5 \omega^2}\left( \omega^3 + (k_2+k_3)(k_2 k_3- \omega^2) + \omega(k_2^2 + k_3^2)\right)
\times \left(1 + {\cal O}(\tau) \right) \\
{J_{1b}} &=&   {\rm Im}\,\left\{  \int_{\tau}^\infty d\tau_1 {\tau_1}^{-4} \,
\frac{d}{d\tau_1} u_{\frac{5}{2}}(x_2 \tau_1) \, \frac{d}{d\tau_1} u_{\frac{5}{2}}(x_3 \tau_1) \,
 \left( \tau_1 \frac{d}{d\tau_1}\right)u_{\frac{3}{2}}(\tau_1) \right\} \cr\cr
&=& \frac{k_2^2 k_3^2}{9 k_1^3 \omega^3}
\left(  \omega^2 + \omega(k_2+k_3)+2 k_2 k_3 \right)  \times \left(1 + {\cal O}(\tau) \right)  \\
{J_{2a}}&=&     {\rm Im}\,\left\{ \int_{\tau}^\infty d\tau_1   {\tau_1}^{-4} \,
u_{\frac{5}{2}}(x_2 \tau_1) u_{\frac{5}{2}}(x_3 \tau_1) \,  u_{\frac{3}{2}}(\tau_1) \right\} \cr
&=&  -\frac{1}{3}\left(\gamma + \ln[\tau \omega/k_1] \right)+ \frac{1}{9 k_1^3 \omega^2}
\left( k_2^2 k_3^2(k_1+\omega) -3 \omega k_2 k_3(\omega^2 - k_1 \omega + k_1^2)+3 k_1 \omega^3(2 k_1 - \omega)\right)\times \left(1 + {\cal O}(\tau) \right) \\
{J_{2b}}& =&   {\rm Im}\,\left\{    \int_{\tau}^\infty d\tau_1   {\tau_1}^{-4} \,
u_{\frac{5}{2}}(x_2 \tau_1) u_{\frac{5}{2}}(x_3 \tau_1) \, \left( \tau_1 \frac{d}{d\tau_1}\right)u_{\frac{3}{2}}(\tau_1) \right\} \cr
&=& - \left(\gamma + \ln[\tau \omega/k_1] \right)+ \frac{1}{9 k_1 \omega^3}\left(9 \omega^4 - 3 \omega^2(k_2^2 + 4 k_2 k_3 + k_3^2)+ 3 \omega k_1 k_2 k_3 - 2 k_2^2 k_3^2\right)\times \left(1 + {\cal O}(\tau) \right) \\
J_{3a} &=&  {\rm Im}\,\left\{   \int_{\tau}^\infty d\tau_1  {\tau_1}^{-4} \,
u_{\frac{5}{2}}(x_2 \tau_1) \left( \tau_1 \frac{d}{d\tau_1}u_{\frac{5}{2}}(x_3 \tau_1)  \right) u_{\frac{3}{2}}(\tau_1) \right\} \cr
&=& \frac{ k_3^2}{9 k_1^3 \omega^3} \left(
3 \omega^4 -3 (k_2 + k_3)\omega^3 + (k_2^2 + 3 k_3^2)\omega^2 + k_2(k_2^2 + k_2 k_3 + 3 k_3^2)\omega + 2 k_2^2 k_3(k_2+k_3)\right) \times \left(1 + {\cal O}(\tau) \right)\\
J_{3b} &=&  {\rm Im}\,\left\{   \int_{\tau}^\infty d\tau_1  {\tau_1}^{-4} \,
u_{\frac{5}{2}}(x_3 \tau_1) \left( \tau_1 \frac{d}{d\tau_1}u_{\frac{5}{2}}(x_2 \tau_1)  \right) u_{\frac{3}{2}}(\tau_1) \right\} \cr
&=& \frac{ k_2^2}{9 k_1^3 \omega^3} \left(
3 \omega^4 -3 (k_2 + k_3)\omega^3 + (k_3^2 + 3 k_2^2)\omega^2 + k_3(k_3^2 + k_2 k_3 + 3 k_2^2)\omega + 2 k_3^2 k_2(k_2+k_3)\right)\times \left(1 + {\cal O}(\tau) \right)
\end{eqnarray}
Hence, $K_1 = -4 k_1 k_2 k_3 (J_{1a}+J_{1b})$,  $K_2 = 2 (k_2^2 k_3^2/k_1)(J_{2a}+J_{2b})$, and $K_{3a,b} = 2 (k_2^2 k_3^2/k_1) J_{3a,b}$.  In all cases we have assumed $\tau \ll 1$. 
 
%%%%%%%%%%%%%%%%%%%%%%%%%%%%%%%%%%%%%%%%%%%%%%%%%%%%%%%%%%%%
\acknowledgments
This work was supported in part by NSF PHY-1068027 at Dartmouth College. LM would like to thank Marco Peloso, Neil Barnaby and Ryo Namba for useful discussions and criticism. We would like to thank Marc Kamionkowski for useful conversations.

\vfill  

%%%%%%%%%%%%%%%%%%%%%%%%%%%%%%%%%%%%%%%%%%%%%%%%%%%%%%%%%%%%

\end{document}